\documentclass[preprint]{JHEP3new} 

\JHEPspecialurl{http://jhep.sissa.it/JOURNAL/JHEP3.tar.gz}

\usepackage{epsfig,multicol,bbm,rotating}

\newcommand\fverb{\setbox\pippobox=\hbox\bgroup\verb}
\newcommand\fverbdo{\egroup\medskip\noindent%
            \fbox{\unhbox\pippobox}\ }
\newcommand\fverbit{\egroup\item[\fbox{\unhbox\pippobox}]}
\newbox\pippobox

\def\ifm#1{\relax\ifmmode#1\else$#1$\fi}
    
\def\x{\ifm{\times}}   \def\pt#1,#2,{\ifm{#1\x10^{#2}}}
\def\up#1{\ifm{^{#1}}}     \def\plm{\ifm{\,\pm}\,}

\def\rmk{\rm\kern.5mm }

   \def\minus{$-$}

\def\figb#1;#2;{\parbox{#2cm}{\epsfig{file=#1.eps,width=#2cm}}}

\newcommand{\bea}{\begin{eqnarray}}
\newcommand{\eea}{\end{eqnarray}}

\newcommand{\be}{\begin{equation}}
\newcommand{\ee}{\end{equation}}
\newcommand{\beq}{\begin{equation}}
\newcommand{\eeq}{\end{equation}}
\newcommand{\ba}{\begin{array}}
\newcommand{\ea}{\end{array}}
\newcommand{\beqa}{\begin{eqnarray}}
\newcommand{\eeqa}{\end{eqnarray}}

\def\Im{{\rm Im}}

\newcommand{\Fig}[1]{Fig.~\ref{#1}}

\newcommand{\Tab}[1]{Table~\ref{#1}}

\newcommand{\etal}{et al.}

\newcommand{\BR}[1]{\ensuremath{{\rm BR}(#1)}}

\newcommand{\SN}[2]{\ensuremath{#1\times10^{#2}}}

\newcommand{\Vud}{\ensuremath{|V_{ud}|}}
\newcommand{\Vus}{\ensuremath{|V_{us}|}}
\newcommand{\Vusf}{\ensuremath{|V_{us}|f_+(0)}}

\newcommand{\fp}{f_+(0)}

\newcommand{\mev}{{\rm MeV}}
\newcommand{\text}{\rm}

\catcode`@=11 
\newdimen\z@ \z@=0pt 
\newskip\z@skip \z@skip=0pt plus0pt minus0pt
\def\m@th{\mathsurround=\z@}
\def\ialign{\everycr{}\tabskip\z@skip\halign} 
\def\eqalign#1{\null\,\vcenter{\openup\jot\m@th
  \ialign{\strut\hfil$\displaystyle{##}$&$\displaystyle{{}##}$\hfil
      \crcr#1\crcr}}\,}
\catcode`@=12 

\title{Precision tests of Standard Model  with leptonic and semileptonic
        kaon decays}
\author{M. Antonelli \footnote{most of this material is done in conjunction
 with \cite{flavia}}
 \small{INFN, Laboratori Nazionali di Frascati, Via E.Fermi 40, I-00044 Frascati, Italy}
}

 \received{\today}      
\revised{\today}
\accepted{\today}       

\abstract{
Till the middle of 2004 it appeared that the unitarity relation 
$\vert V_{ud}\vert^2 +\vert V_{us}\vert^2 +
\vert V_{ub}\vert^2 = 1$ might not hold at the $2.3\sigma$
level.  At that time, however,  $\vert V_{us}\vert$ was inferred from  old experimental
data.   Since then, a large experimental and theoretical effort has been invested leading to
a removal of the problem. Thanks to the new and improved measurements  by BNL-E865, KLOE,
KTeV, ISTRA+ and NA48,  the $K_{\ell3}$ decay rate moved up so that  $|V_{us}|$ is now  consistent with unitarity. On the theory side, much progress has been made in order to tame  the systematic uncertainties related to
the computation  of the $K_{\ell3}$ form factors.

This joint progress allowed to assess the validity of the CKM unitarity
relation at the level of
less than $1\%$.
   Recent measurements of kaon decays contributing to the determination of
  \Vus\, \Vus/\Vud\ are summarized, and up-to-date evaluations of \Vusf\ and \Vus\
   are presented. 
In addition, we discuss the sensitivity of  $K_{\ell3}$ and $K_{\ell2}$ decays
 to various scenarios of physics beyond Standard Model.
}

\keywords{Vus, CKM, kaon}

\begin{document}
\section{Introduction}
 I report on precise tests of the Standard Model (SM) with
 kaon decays using world data.
From the experimental information on down- to up-quark transitions
(such as $d\to u$, $s\to u$ and $b\to u$), we access the effective
 dimesion-six operators of the form, $\overline D\,\Gamma_1 U \bar\ell\,\Gamma_2\,\nu$, with
 $D$ ($U$) being a generic ``{\it down}" (``{\it up}") flavor, and
  $\ell=e,\,\mu,\,\tau$.
 Their effective coupling are parametrized as the SM contribution
 $G_F^2\,\vert V_{UD}\vert^2$, plus a possible new physics
 terms, $G_F^2\,\epsilon_{NP}$. Since  the dimension-six operators
 are not protected
 by gauge invariance the possible  effects of non-decoupling are
 proportional to
 $(1+ M^2_W/\Lambda^2_{NP}$).
  The effects of these non-standard contributions
 cannot be very large, but are possibly detectable in high-precision
 experiments.

 A convenient strategy
 to measure these effects against the SM parameters, $G_F^2$ and
 $\vert V_{UD}\vert$,  rely on the Cabibbo universality hypothesis
 (or unitarity constraint):
\begin{equation}
G^2_{CKM}=G^2_\mu\  \left(\mbox{or}\ 
|V_{ud}|^2+|V_{us}|^2+|V_{ub}|^2=1\ \mbox{and}\ G_F\equiv G_\mu \right) ,
\label{eq:unitarity}
\end{equation}
 where $G^2_{CKM}=G_F^2\left( |V_{ud}|^2 + |V_{us}|^2 +|V_{ub}|^2\right)$, and
 $G_\mu = 1.166371 (6) \times 10^{-5} {\rm GeV}^{-2}$, as extracted from the
 accurate measurement of the  muon lifetime~\cite{mulan}.

 I report on the progress related to the verification of the unitarity
 relation~(\ref{eq:unitarity}). As we shall see the current
 accuracy of the CKM unitarity relation~(\ref{eq:unitarity}),
 is at the $0.1\%$ level, becoming an important
 constraint to the model builders of scenarios beyond SM physics.
$K_{\ell3}$ and $K_{\mu2}$ decays offer possibly the cleanest way
 to test $us$ transitions. ${ud}$ transitions are precisely measured in
 superallowed nuclear $\beta$-decays. The most recent determination
 of $V_{ud}$ is $|V_{ud}|$=0.97418\plm0.00026, \cite{t&h}.

 This report is organized as follows. The phenomenological
 farmework needed to describe  $K_{\ell3}$ and $K_{\mu2}$ decays is
 briefly recapitulated in Section~\ref{sec:master}.  Section~\ref{sec:data}
 is dedicated to the combination of the experimental data.
 The results and  interpretations are
 presented in Section~\ref{sec:results}.

\subsection{\boldmath{$K_{\ell 2}$} and \boldmath{$K_{\ell 3}$} phenomenology}
\label{sec:master}

For $K_{\ell 2}$ ($\pi_{\ell 2}$) amplitudes, we introduce the following QCD parameters
\begin{equation}
\langle0|\bar{s}\gamma_{\mu}\gamma_{5}u|K^{+}\left(  p\right)  \rangle
=i\sqrt{2}f_{K}\,p_{\mu},\;\;\langle0|\bar{s}\gamma_{5}u|K^{+}\rangle=-i\sqrt
{2}f_{K}\frac{m_{K}^{2}}{m_{s}+m_{u}}\;, \label{MEkl2}%
\end{equation}

For $K_{\ell 3}$ amplitudes, we define the following form factors
\be\eqalign{
\langle\pi^{+}\left(  k\right)  |\bar{s}\gamma^{\mu}u|K^{0}\left(  p\right)
\rangle&=\frac{1}{\sqrt{2}}\left( (p+k) ^\mu f_+^{K^0\pi^+}(t) +(p-k) ^\mu f_-^{K^0\pi^+}(t) \right)\cr
f_-^{K\pi}(t)&=\frac{m^2_K-m^2_\pi}{t}\left(f_0^{K\pi}(t)  -f_+^{K\pi}(t)\right)\cr}
\label{Eq4}
\ee
where $t=(p-k)^2$ and
\begin{equation}
\langle\pi^+\left(  k\right)|\bar{s}u|K^0\left(  p\right)\rangle=-\frac{M_{K}^{2}-M_{\pi}^{2}}{\sqrt
{2}\left(  m_{s}-m_{u}\right)  }f_{0}\left(  t\right)  \; \label{SP1}%
\end{equation}

The SM gives the following relations for the $K_{\ell3}$ and
$K_{\ell2}$ decay rates:
\bea
\label{eq:Mkl3}
\Gamma(K_{\ell 3(\gamma)}) &=&
{ G_\mu^2 M_K^5 \over 192 \pi^3} C_K
  S_{\rm ew}\,|V_{us}|^2 f_+(0)^2\,
I_K^\ell(\lambda_{+,0})\,\left(1 + \delta^{K}_{SU(2)}+\delta^{K \ell}_{\rm
em}\right)^2\,\quad ,\\
\frac{\Gamma(K^{\pm}_{\ell 2(\gamma)})}{\Gamma(\pi^{\pm}_{\ell 2(\gamma)})} &=&
\large\left|\frac{V_{us}}{V_{ud}}\large\right|^2\frac{f^2_K m_K}
{f^2_\pi m_\pi}\left(\frac{1-m^2_\ell/m_K^2}{1-m^2_\ell/m_\pi^2}\right)
\times\left(1+\delta_{\rm
em}\right)\quad ,\label{eq:Mkl2}
\eea
where  $C_{K}=1$  ($1/2$) for the neutral (charged) kaon decay.
$I_K^\ell(\lambda_{+,0})$ is the phase space integral which also includes the
form factors
parameterized by $\lambda_{+,\,0}$.  The universal short-distance electromagnetic
correction, $S_{\rm ew}=1.0232(3)$,  has been computed at  $\mu=M_\rho$ in ref.~\cite{Sirlin:1981ie}, while
 the long-distance electromagnetic corrections,
 $\delta_{\rm em}=0.9930(35)$\cite{ciriglianokl2} and $\delta^{K \ell}_{\rm
em}$, as well as  the  isospin-breaking ones,
$\delta^{K}_{SU(2)}$, have been recently revised in ref.~\cite{Cirigliano}
(see table~\ref{tab:iso-brk}).

\begin{table}[h]
\setlength{\tabcolsep}{3.8pt}
\centering
\begin{tabular}{c||c||c|}
& $\delta^K_{SU(2)} (\%)$
& $\delta^{K \ell}_{\rm em}(\%) $   \\
\hline
$K^{0}_{e 3}$   &  0        &  +0.57(15\\
$K^{+}_{e3}$    & 2.36(22)  &  +0.08(15)\\
$K^{0}_{\mu 3}$ &  0        &  +0.80(15)\\
$K^{+}_{\mu 3}$ & 2.36(22)  &  +0.05(15)\\
\end{tabular}
\caption{Summary of the isospin-breaking
factors~\cite{neufeld,Cirigliano}}
\label{tab:iso-brk}
\end{table}

with correlations for $\delta^{K \ell}_{\rm em}(\%)$:
\be
\left(
\begin{array}{cccc}
  1.  &  0.11 &  0.78& -0.12\\
      &   1.  & -0.12&  0.78\\
      &       &    1.&  0.11\\
      &       &      &     1.\\

\end{array}
\right)
\ee
The remaining quantities, $f_{+}(0)$,  the vector form factor at 
 zero momentum transfer [$q^2=(p_K-p_\pi)^2= 0$], and $f_K/f_\pi$, the
 ratio of the kaon and pion decay constants contains the non-perturbative
 QCD information on the flavor SU(3)  breaking effects arising in the
 relevant hadronic matrix element. For more details see Sec. \ref{sec:fofzero}.

\subsection{Form Factors Parameterizations: Lattice QCD,  ChPt and Dispersion relations}
\label{sec:ffpara}
To determine  $V_{us}$, we have to determine the integral over phase space of the Dalitz density, which
depends on the form of $f_{+,\,0}(t)$, the form factors (FF) in eq.~(\ref{Eq4}).
To reduce uncertainties, it would be convenient to have theoretical
information on their $t$ dependence. Our theorical knowledge is still poor.
ChPt and Lattice QCD could be very useful but predictivity of ChPt
is limited by our knowledge of  Low Energy Constants(LECs),  while  lattice calculations still have large uncertainty.

For this reason, present estimates of the phase space integral mainly rely on
measurements. In the physical region,
it is a good approximation  to use a quadratic $t$ dependence of the FFs such as
\begin{equation}
  \tilde{f}_{+,\,0}(t) \equiv\frac{{f}_{+,\,0}(t)}{{f}_{+,\,0}(0)} = 1 + \lambda'_{+,\,0}~\frac{t}{m^2}+\frac{1}{2}\;\lambda''_{+,\,0}\,\left(\frac{t}{m^2}\right)^2
  \label{eq:ff2}
\end{equation}
Note that $t=(p_K-p_\pi)^2=m_K^2+m_\pi^2-2m_KE_\pi$, therefore the FFs depends only on $E_\pi$.
The FF parameters can thus be obtained from a fit to the pion spectrum which is of the form
$g(E_\pi)\x\tilde f(E_\pi)^2$. Unfortunately $t$ is maximum for $E_\pi$=0, where $g(E_\pi)$ vanishes.
Still, experimental information about $\tilde{f}_{+}$ are quite accurate
and a pole parametrization, $\lambda''_{+}\sim 2(\lambda'_{+})^2$,
looks confirmed from present data (see later).
 Theory would also support this hypothesis. The situation of
 $\tilde{f}_{0}$ instead is completely open. The main problem is that
  $\lambda''_{0}$ cannot be detected from the data and we can not
  discriminate among
  different assumptions such as linear, pole or
 quadratic fits. In turn,  these model ambiguities induce a
 systematics uncertainty for \Vus\, even though data for partial rates
 by itself are very accurate. For this reason, hints from theory are welcome
 and  it is advisable to use model-independent
parametrisations as possible.

The vector and scalar form factors $f_{+/0}(t)$ in eq.~(\ref{Eq4})
are analytic functions in the complex
$t$--plane, except for a cut along the positive real axis, starting at the
first physical threshold $t_{th\,1} = (M_K+M_\pi)^2$, where their imaginary
parts develop discontinuities.  They are real for $s<t_{th\,1}$.

Cauchy's theorem implies that $f_{+/0}(t)$ can be written as a dispersive integral along the physical cut
\begin{equation}
\label{dis}
f(t) \; = \; \frac{1}{\pi} \int\limits^\infty_{t_{th\,1}}\!\! ds'\,
\frac{\Im f(s')}{(s'-t-i0)} + {\rm subtractions} \,.
\end{equation}
where  all possible on-shell intermediate states contribute.
A number of subtractions are needed to make the integral convergent.
 Particularly appealing is an improved dispersion relation recently
 proposed in ref.~\cite{stern}, which reads
\def\ct{{\rm CT}}
\be
\tilde f_0(t)=\left(\tilde f_0(t_{CT})
\,e^{\left(\frac{t}{t_{\ct}}-1\right)H(t)}\right)^{\frac{t}{t_{\ct}}},\kern3mm
H(t)=\frac{t_{\ct}^2}{\pi}\int^\infty_{t_{\ct}}\frac{dx}{x}
\frac{\phi(x)}{\left(x-t_{\ct}\right)\left(x-t-i\epsilon\right)}
\label{eq:stern}
\ee
Here $\phi(x)$, the phase of $f_0(t)$, is taken from $K\pi$ scattering.
This dispersive form has been solved in terms of
 the subtractions at $t=0$
(where by default, $\tilde f_0(0)\equiv 1$ is known) and at
$t_{CT}=(m_K^2-m_\pi^2)$, where the
QCD soft-pion theorem, known as  Callan-Treiman relation, implies
\be
\tilde f_0(t\equiv(m_K^2-m_\pi^2))=\frac{f_K}{f_\pi}\frac{1}{f_+(0)} + \,(3.5\pm8.0\,10^{-3}\cite{leut})
\ee
By exploiting this form on the spectrum of $K_{\ell3}$ data, we have to estimate
one unknown parameter ($\tilde
 f_0(t\equiv(m_K-m_\pi)^2)$ or $f_K/f_\pi/f_+(0)$), whereas, eq.~(\ref{eq:stern})
 within its theoretical uncertainty non-trivially contraints
the coefficients of the Taylor's expansion in eq.~(\ref{eq:ff2})(for
axample for the first two derivatives, we have from ref.~\cite{stern} $\lambda''_0=
(\lambda'_0)^2  -2m^4_\pi/t_{CT} G'(0) =  (\lambda'_0)^2+ (4.16 \pm 0.50)\times
10^{-4}$). By the experiemtal information of $\tilde
 f_0(t\equiv(m_K-m_\pi)^2)$, Callan-Treiman theorem allows for a consistency
 checks with lattice QCD, where $f_K/f_\pi/f_+(0)$ can be estimated (see section\ref{sec:CTtest}).

\section{Data Analysis}
\label{sec:data}
We perform fits to world data on the BRs and lifetimes for the
$K_L$ and $K^\pm$, with the constraint that BRs add to unity\cite{flavia}.
This is the most satisfactory way of making use
of the new measurements. 
\subsection{$K_L$   leading branching ratios and $\tau_L$ }
\label{sec:KL}
Numerous measurements of the principal $K_L$ BRs, or of various ratios
of these BRs, have been published recently. For the purposes of evaluating
\Vusf, these data can be used in a PDG-like fit to the $K_L$ BRs and lifetime,
so all such measurements are of interest.

KTeV has measured five ratios of the six main $K_L$ BRs~\cite{KTeV+04:BR}.
The six channels
involved account for more than 99.9\% of the $K_L$ width and KTeV combines the
five measured ratios to extract the six BRs. We use the five measured ratios
in our analysis:
$\BR{K_{\mu3}/K_{e3}} = 0.6640(26)$,
$\BR{\pi^+\pi^-\pi^0/K_{e3}} = 0.3078(18)$,
$\BR{\pi^+\pi^-/K_{e3}} = 0.004856(28)$,
$\BR{3\pi^0/K_{e3}} = 0.4782(55)$, and
$\BR{2\pi^0/3\pi^0} = 0.004446(25)$. The errors on these measurements are
correlated; this is taken into account in our fit.

NA48 has measured the ratio of the BR for $K_{e3}$ decays to the sum of BRs
for all decays to two tracks, giving
$\BR{K_{e3}}/(1-\BR{3\pi^0}) = 0.4978(35)$ \cite{NA48+04:BR}. From a
separate measurement of \BR{K_L\to3\pi^0}/\BR{K_S\to2\pi^0}, NA48
obtains $\BR{3\pi^0}/\tau_L = 3.795(58)$ $\mu$s\up{-1} \cite{Lit04:ICHEP}.

Using $\phi\to K_L K_S$ decays in which the $K_S$ decays to $\pi^+\pi^-$,
providing normalization, KLOE has directly measured the BRs for the four
main $K_L$ decay channels \cite{KLOE+06:BR}.
The errors on the KLOE BR values are dominated
by the uncertainty on the $K_L$ lifetime $\tau_L$; since the dependence of
the geometrical efficiency on $\tau_L$ is known, KLOE can solve for $\tau_L$
by imposing $\sum_x \BR{K_L\to x} = 1$ (using previous averages for the minor
BRs), thereby greatly reducing the uncertainties on the BR values obtained.
Our fit makes use of the KLOE BR values before application of this constraint:
\BR{K_{e3}} = 0.4049(21),
\BR{K_{\mu3}} = 0.2726(16),
\BR{K_{e3}} = 0.2018(24), and
\BR{K_{e3}} = 0.1276(15).
The dependence of these values on $\tau_L$ and the correlations between the
errors  are taken into account.
KLOE has also measured $\tau_L$ directly, by fitting the proper decay time
distribution for $K_L\to3\pi^0$ events, for which the reconstruction
efficiency is high and uniform over a fiducial volume of $\sim$$0.4\lambda_L$.
They obtain $\tau_L=50.92(30)$~ns \cite{KLOE+05:tauL}.

There are also two recent measurements of \BR{\pi^+\pi^-/K_{\ell3}},
in addition to the KTeV measurement of \BR{\pi^+\pi^-/K_{e3}} discussed above.
KLOE obtains \BR{\pi^+\pi^-/K_{\mu3}} =\break \SN{7.275(68)}{-3} \cite{KLOE+06:KLpp},
while NA48 obtains \BR{\pi^+\pi^-/K_{e3}} = \SN{4.826(27)}{-3}
\cite{NA48+06:KLpp}. All measurements are fully inclusive of inner
bremsstrahlung. The KLOE measurement is fully inclusive of the direct-emission
(DE) component, DE contributes negligibly to the KTeV measurement, and a
residual DE contribution of 0.19\% has been subtracted from the NA48 value
to obtain the number quoted above. For consistency, in our fit,
a DE contribution of 1.52(7)\% is added to the KTeV and NA48 values.
Our fit result for \BR{\pi^+\pi^-} is then understood to be DE inclusive.

In addition to the 14 recent measurements listed above, our fit for the
seven largest $K_L$ BRs and lifetime uses four of the remaining five
inputs to the 2006 PDG fit and the constraint that the seven BRs add to unity.
The results are given in \Tab{tab:KLBR}.

\begin{table}
\begin{center}
\begin{tabular}{l|c|r}
Parameter & Value & $S$ \\
\hline
\BR{K_{e3}} & 0.40563(74) & 1.1 \\
\BR{K_{\mu3}} & 0.27047(71) & 1.1 \\
\BR{3\pi^0} & 0.19507(86) & 1.2 \\
\BR{\pi^+\pi^-\pi^0} & 0.12542(57) & 1.1 \\
\BR{\pi^+\pi^-} & \SN{1.9966(67)}{-3} & 1.1 \\
\BR{2\pi^0} & \SN{8.644(42)}{-4} & 1.3 \\
\BR{\gamma\gamma} & \SN{5.470(40)}{-4} & 1.1 \\
$\tau_L$ & 51.173(200)~ns & 1.1 \\
\end{tabular}
\end{center}
\vskip 0.3cm
\caption{\label{tab:KLBR}
Results of fit to $K_L$ BRs and lifetime}
\end{table}
The evolution of the average values of the BRs for
$K_{L\ell3}$ decays and for
the important normalization channels is shown in \Fig{fig:kpmavg}.
\begin{figure}[h]
\begin{center}
\includegraphics[width=0.9\textwidth]{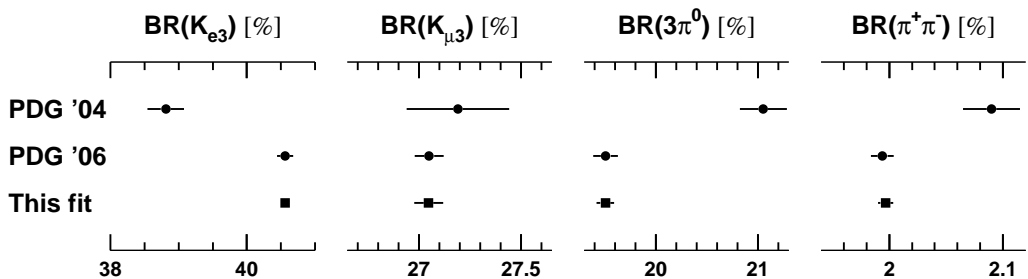}
\end{center}
\caption{\label{fig:klavg}
Evolution of average values for main $K_L$ BRs.}
\end{figure}

Our fit gives $\chi^2/{\rm ndf} = 20.2/11$ (4.3\%), while the 2006
PDG fit gives $\chi^2/{\rm ndf} = 14.8/10$ (14.0\%).
The differences between the output values from our fit and the
2006 PDG fit are minor.
The poorer value of $\chi^2/{\rm ndf}$ for our fit can be traced
to contrast between the KLOE value for \BR{3\pi^0} and the
other inputs involving \BR{3\pi^0} and \BR{\pi^0\pi^0}---in
particular, the PDG ETAFIT value for \BR{\pi^0\pi^0/\pi^+\pi^-}.
The treatment of the correlated KTeV and KLOE measurements in the
2006 PDG fit gives rise to large
scale factors for \BR{K_{e3}} and \BR{3\pi^0};
in our fit, the scale factors are more uniform. As a result,
our value for \BR{K_{e3}} has a significantly smaller uncertainty
than does the 2006 PDG value.

\subsection{$K_S$  leading branching ratios and $\tau_S$}
\label{sec:KS}
KLOE has published \cite{KLOE+06:KSe3}
a measurement of \BR{K_S\to\pi e\nu} that is precise enough
to contribute meaningfully to the evaluation of \Vusf.
The quantity directly measured is \BR{\pi e\nu}/\BR{\pi^+\pi^-}. Together
with the  published KLOE value
\BR{\pi^+\pi^-}/\BR{\pi^0\pi^0} = 2.2459(54),
the constraint that the $K_S$ BRs must add to unity, and the assumption of
universal lepton couplings, this completely determines the $K_S$ BRs for
$\pi^+\pi^-$, $\pi^0\pi^0$, $K_{e3}$, and $K_{\mu3}$ decays
\cite{KLOE+06:KSpp}. In particular, $\BR{K_S\to\pi e\nu} = \SN{7.046(91)}{-4}$.

NA48 has recently measured the ratio
$\Gamma(K_S \to \pi e \nu)/\Gamma(K_L \to \pi e \nu) = 0.993(26)(22)$
 \cite{NA48:KSe3}.
The best way to include this measurement in our analysis would
be via a combined fit to $K_S$ and $K_L$ branching ratio and lifetime
measurements. Indeed, such a fit would be useful in properly
accounting for correlations between $K_S$ and $K_L$ modes introduced
with the preliminary NA48 measurement of $\Gamma(K_L\to 3\pi^0)$, and
more importantly, via the PDG ETAFIT result, which we use in the
fit to $K_L$ branching ratios. At the moment, however, we fit
$K_S$ and $K_L$ data separately. NA48 quotes
$\BR{K_S\to\pi e\nu} = \SN{7.046(180)(160)}{-4}$;
averaging this with the KLOE result gives
$\BR{K_S\to\pi e\nu} = \SN{7.046(84)}{-4}$,
 improving the accuracy on this BR by about 10\%. 

For $\tau_{K_S}$ we use \SN{0.8958}{-10}~s, where this is the non-$CPT$
constrained fit value from the PDG, and is dominated by the 2002 NA48
and 2003 KTeV measurements.

\subsection{$K^\pm$ leading branching ratios and $\tau^\pm$}
There are several new results providing information on $K^\pm_{\ell3}$
rates. These results are mostly preliminary and have not been included
in previous averages.

NA48/2 has recently  published measurements of the three ratios
\BR{K_{e3}/\pi\pi^0}, \BR{K_{\mu3}/\pi\pi^0}, and
\BR{K_{\mu3}/K_{e3}} \cite{NA48+07:BR}.
These measurements are not independent; in our fit, we use the values
$\BR{K_{e3}/\pi\pi^0} = 0.2470(10)$ and
$\BR{K_{\mu3}/\pi\pi^0} = 0.1637(7)$ and take their correlation
into account.
ISTRA+ has also updated its preliminary value for $\BR{K_{e3}/\pi\pi^0}$.
They now quote $\BR{K_{e3}/\pi\pi^0} = 0.2449(16)$\cite{Rom06:ke3}.

KLOE has measured the absolute BRs for the
$K_{e3}$ and $K_{\mu3}$ decays
\cite{KLOE:kl3pm} and a very precise measuremet of 
 $\BR{K_{\mu2}}$\cite{KLOE05:kl2}.
In $\phi\to K^+ K^-$ events, $K^+$ decays into $\mu\nu$ or $\pi\pi^0$
are used to tag a $K^-$ beam, and vice versa. KLOE performs four
separate measurements for each $K_{\ell3}$ BR, corresponding to the
different combinations of kaon charge and tagging decay.
The final averages are $\BR{K_{e3}} = 4.965(53)\%$ and
$\BR{K_{\mu3}} = 3.233(39)\%$.

Very recently KLOE has also measured the absolute
branching ratio for the $\pi\pi^0$ decay with 0.5\% accuracy.
The KLOE preliminary result, is  $\BR{\pi\pi^0}=0.20658(112)$\cite{KLOE:pipo}.

 Our fit takes into account the correlation between these values, as
 well as their dependence on the $K^\pm$ lifetime.
 The world average value for $\tau_\pm$ is nominally
 quite precise; the 2006 PDG quotes $\tau_\pm = 12.385(25)$~ns.
 However, the error is scaled by 2.1; the confidence level for the
 average is 0.17\%. It is important to confirm the value of $\tau_\pm$.
 The two new measurements from KLOE,
 $\tau_\pm = 12.364(31)(31)$~ns  and
 $\tau_\pm = 12.337(30)(20)$~ns\cite{KLOE:taupm} with correlation 34\%,
 agree with the PDG average.

 Our fit for the six largest $K^\pm$ BRs and lifetime makes use of the
 results cited above,
 plus the data used in the 2006 PDG fit, except for the
 Chiang '72 measurements
for a total of 26 measurements.
The six BRs are constrained to add to unity.
The results are shown in \Tab{tab:KpmBR}.
\begin{table}
\begin{center}
\begin{tabular}{l|c|r}
Parameter & Value & $S$ \\
\hline
\BR{K_{\mu2}}      & 63.569(113)\%   & 1.1 \\
\BR{\pi\pi^0}      & 20.644(80)\%   & 1.1 \\
\BR{\pi\pi\pi}     &  5.5953(308)\%  & 1.0 \\
\BR{K_{e3}}        &  5.0780(258)\%    & 1.2 \\
\BR{K_{\mu3}}      &  3.3650(271)\%  & 1.7 \\
\BR{\pi\pi^0\pi^0} &  1.7495(261)\%  & 1.1 \\
$\tau_\pm$         & 12.3840(193)~ns & 1.7 \\
\end{tabular}
\end{center}
\vskip 0.3cm
\caption{\label{tab:KpmBR}
Results of fit to $K^\pm$ BRs and lifetime}

\end{table}
\begin{figure}[h]
\begin{center}
\includegraphics[width=0.9\textwidth]{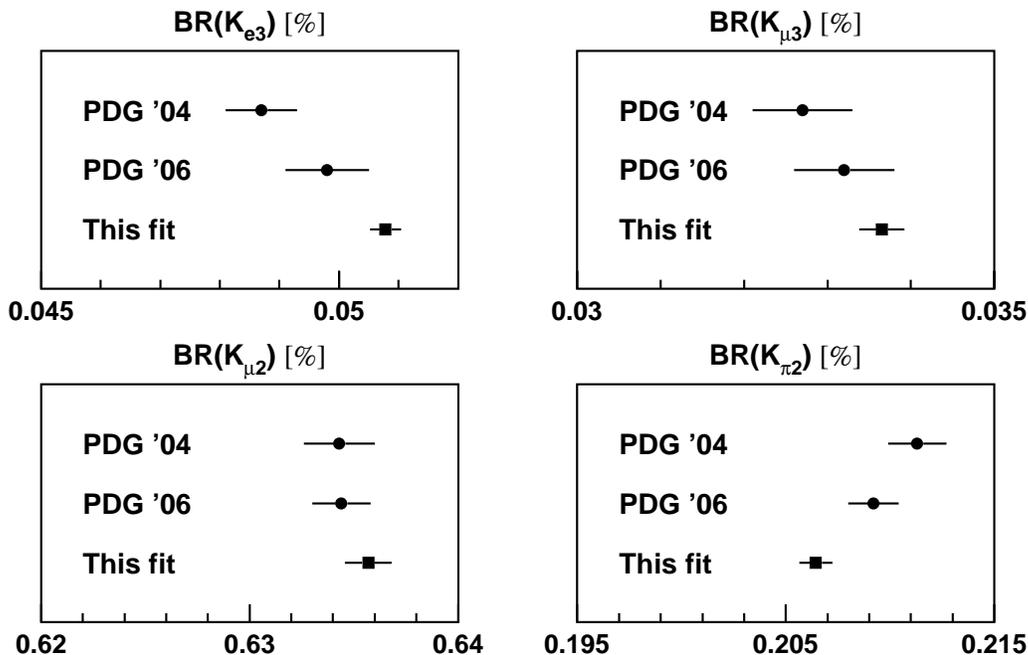}
\end{center}
\caption{\label{fig:kpmavg}
Evolution of average values for main $K^\pm$ BRs.}
\end{figure}

The fit quality is poor, with $\chi^2/{\rm ndf} = 42/20$ (0.31\%).
However, when the five older measurements of $\tau_\pm$ are replaced
by their PDG average with scaled error,  $\chi^2/{\rm ndf}$ improves
to 24.3/16 (8.4\%), with no significant changes in the results.

Both the significant evolution of the average values of the $K_{\ell3}$
BRs and the effect of the correlations with \BR{\pi\pi^0} are evident
in \Fig{fig:kpmavg}.

\subsection{Measurement of BR($K_{e2})$/BR($K_{\mu2}$)}
Experimental knowledge of $K_{e2}/K_{\mu2}$ has been poor so far.
 The current world average
of $R_K = \BR{K_{e2}}/\BR{K_{\mu2}}= (2.45 \pm 0.11) \times 10^{-5}$ dates back to three experiments
of the 1970s~\cite{bib:pdg} and has a precision of about 5\%.
The three new preliminary measurements were reported by NA48/2 and KLOE
 (see Tab.~\ref{tab:ke2kmu2}):
A preliminary result of NA48/2, based on about 4000 $K_{e2}$ events from the 2003
 data set, was presented in 2005~\cite{bib:Ke2_2003}.
Another preliminary result, based on also about 4000 events, recorded in a minimum bias
 run period in 2004, was shown at KAON07\cite{bib:Ke2_2004}.
Both results have independent statistics and are also independent in the systematic uncertainties,
as the systematics are either of statistical nature (as e.g.\ trigger efficiencies) or determined in
an independent way.
Another preliminary result, based on about 8000 $K_{e2}$ events, was presented at KAON07
by the KLOE collaboration~\cite{bib:Ke2_KLOE}.
Both, the KLOE and the NA48/2 measurements are inclusive with respect to the final state
 radiation  bremsstrahlung contribution.
The small contribution of $K_{l2\gamma}$ events from direct photon emission from the
 decay vertex was subtracted by each of the experiments.
Combining these new results with the current PDG value yields a current world average of
\begin{equation}
R_K  = ( 2.457 \pm 0.032 ) \times 10^{-5},
\label{eqn:ke2kmu2}
\end{equation}
in very good agreement with the SM expectation and, with a relative error of $1.3\%$,
a factor three more precise than the previous world average.

\begin{table}[t]
  \begin{center}
      \begin{tabular}{lc}
        \hline \hline
                                                  & $R_K$ $[10^{-5}]$  \\ \hline
        PDG 2006~\cite{bib:pdg}                   & $2.45 \pm 0.11$ \\
        NA48/2 prel.\ ('03)~\cite{bib:Ke2_2003}   & $2.416 \pm 0.043 \pm 0.024$ \\
        NA48/2 prel.\ ('04)~\cite{bib:Ke2_2004}   & $2.455 \pm 0.045 \pm 0.041$ \\
        KLOE prel.~\cite{bib:Ke2_KLOE}            & $2.55 \pm 0.05 \pm 0.05$ \\ \hline
        SM prediction                             & $2.477 \pm 0.001$ \\
        \hline \hline
                                                  & \\*[-3mm]
      \end{tabular}
      \caption{Results and prediction for $R_K =  \BR{K_{e2}}/\BR{K_{\mu2}}$.}
      \label{tab:ke2kmu2}
  \end{center}
\end{table}

\subsection{\mathversion{bold}Measurements of $K_{\ell3}$ slopes}

\subsubsection{\mathversion{bold}Vector form factor slopes from  $K_{\ell3}$}

For $K_{e3}$ decays, recent measurements of the quadratic slope parameters
of the vector form factor $({\lambda_+',\lambda_+''})$ are available from
KTeV \cite{KTeV+04:FF},
KLOE \cite{KLOE+06:FF}, ISTRA+ \cite{ISTRA+04:e3FF}, and
NA48 \cite{NA48+04:e3FF}.

We show the results of a fit to the $K_L$ and $K^-$ data in the
first column of \Tab{tab:e3ff}, and  only the $K_L$ data in the
second column. With correlations correctly
taken into account, both fits give good
values of $\chi^2/{\rm ndf}$. The significance of the quadratic
term is $4.2\sigma$ from the fit to all data, and $3.5\sigma$ from
the fit to $K_L$ data only.
\TABLE{
\begin{tabular}{lcc}
\hline\hline
& $K_L$ and $K^-$ data & $K_L$ data only \\
& 4 measurements & 3 measurements \\
& $\chi^2/{\rm ndf} = 5.3/6$ (51\%) &
  $\chi^2/{\rm ndf} = 4.7/4$ (32\%)\\
\hline
\SN{\lambda_+'}{3}
     & $25.15\pm0.87$ & $24.90\pm1.13$ \\
\SN{\lambda_+''}{3}
     & $1.57\pm0.38$ & $1.62\pm0.46$ \\
$\rho(\lambda_+',\lambda+'')$
     & $-0.941$ & $-0.951$ \\
$I(K^0_{e3})$
     & 0.154651(236) & 0.154560(307) \\
$I(K^\pm_{e3})$
     & 0.159005(241) & 0.158912(315) \\
\hline\hline
\end{tabular}
\caption{Average of quadratic fit results for $K_{e3}$ slopes}
\label{tab:e3ff}
}

Including or excluding the $K^-$ slopes
has little impact on the values of $\lambda_+'$ and $\lambda_+''$;
in particular, the values of the phase-space integrals change by just
0.07\%. 

KLOE, KTeV, and NA48 also quote the values shown in \Tab{tab:pole}
for $M_V$ from pole fits to $K_{L\:e3}$ data. The average value of
$M_V$ from all three experiments is
$M_V = 875.3\pm5.4$~MeV with $\chi^2/{\rm ndf} = 1.80/2$.
The three values are quite compatible with each other and
reasonably close to the known value of the $K^{\pm*}(892)$
mass ($891.66 \pm 0.26$~MeV). The values for $\lambda_+'$ and $\lambda_+''$
from expansion of the pole parameterization are qualitatively in
agreement with the average of the quadratic fit results.
More importantly, for the evaluation of the phase-space
integrals, using the average of quadratic or pole fit results gives
values of $I(K^0_{e3})$ that differ by just 0.03\%.
No additional error needs be assigned to account for
differences obtained with quadratic and pole parameterizations for
the vector form-factor slope.
\TABLE{
\begin{tabular}{lc|c}
\hline\hline
Experiment & $M_V$ (MeV) & $\left<M_V\right> = 875.3\pm5.4$ MeV \\
KLOE & $870\pm6\pm7$ &  $\chi^2/{\rm ndf} = 1.80/2$ \\
KTeV & $881.03\pm7.11$ & \SN{\lambda_+'}{3} = 25.42(31) \\
NA48 & $859\pm18$ & $\lambda_+''=2\times\lambda_+'^{\,2}$ \\
& &  $I(K^0_{e3})$ = 0.154695(192) \\
\hline\hline
\end{tabular}
\caption{Pole fit results for $K^0_{e3}$ slopes}
\label{tab:pole}}

\subsubsection{Scalar and Vector form factor slopes from \boldmath{$K_{\ell3}$}}
 For $K_{\mu3}$ decays, recent measurements of the
 slope parameters $({\lambda_+',\lambda_+'',\lambda_0})$ are
 available from  KTeV \cite{KTeV+04:FF},  KLOE \cite{KLOE+07:m3FF},
 ISTRA+ \cite{ISTRA+04:m3FF}, and  NA48 \cite{NA48+06:m3FF}. Note that it is
 not possible, because of correlations, deduce the presence of
 quadratic term in $\tilde f_0(t)$ from the decay spectra.
 For the same reason fits with a linear parametrization give a wrong result
 for the slope $\lambda_0$.

\begin{figure}[h]
\begin{center}
\includegraphics[width=0.95\textwidth]{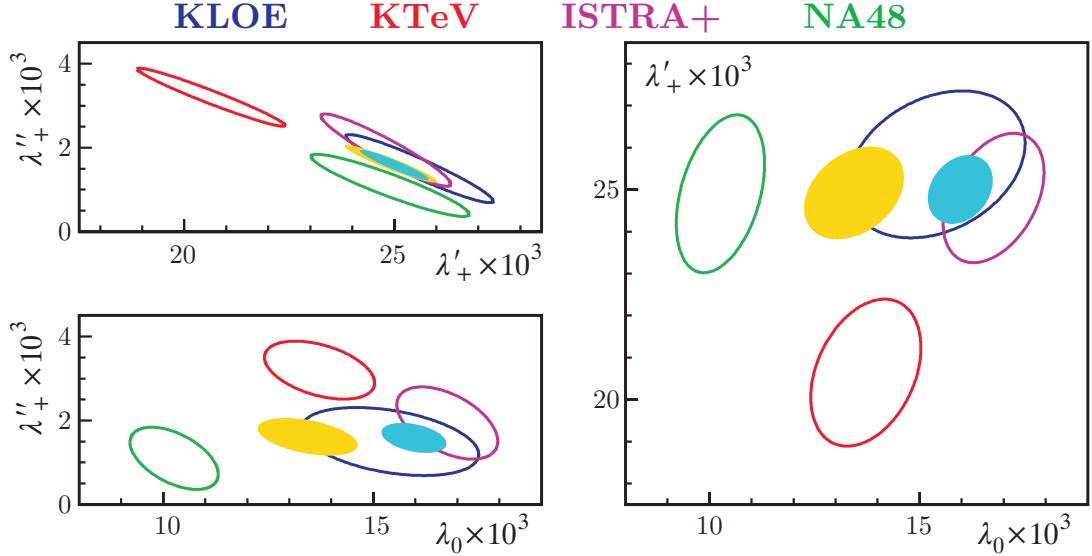}
\end{center}
\caption{\label{fig:FFm3} 1-$\sigma$ contours for $\lambda_+'$, $\lambda_+''$,
  $\lambda_0$ determination from ISTRA+(pink ellipse), KLOE(blue ellipse),
  KTeV(red ellipse), NA48(green ellipse), and world 
 average with(filled yellow ellipse) and without(filled cyan ellipse) the NA48 $K_{\mu3}$ result.}
\end{figure}

 Figure \ref{fig:FFm3} shows the 1-$\sigma$ contours from all
 the experimantal results ($K_{e3}$ and $K_{\mu3}$). It is immediately
 clear from the figure that the new NA48 results are difficult to accommodate.
 Performing the combinaton with and without
 the NA48 results for the $K_{\mu3}$  form-factor slopes included
 we obtain fit probability values of \SN{1}{-6} and 22.3\%
 respectively.
 The results of the combination are listed in \Tab{tab:l3ff}.
\TABLE{
\begin{tabular}{lcc}
\hline\hline
                                 & $K_L$ and $K^-$            & $K_L$ only     \\
\hline
Measurements                     & 16		              & 11	      \\
$\chi^2/{\rm ndf}$               & 54/13 $(7\times 10^{-7})$  & 33/8 $(8\times 10^{-5})$ \\
$\lambda_+'\times 10^3 $         & $24.920\pm1.105$ ($S=1.4$) & $24.011\pm1.544$ ($S=1.5$) \\
$\lambda_+'' \times 10^3 $       & $1.612\pm0.447$ ($S=1.3$)  & $1.974\pm0.622$ ($S=1.6$) \\
$\lambda_0\times 10^3 $          & $13.438\pm1.19$ ($S=1.9$)  & $11.682\pm1.238$ ($S=1.7$) \\
$\rho(\lambda_+',\lambda_+'')$   & $-0.944$                   & $-0.966$       \\
$\rho(\lambda_+',\lambda_0)$     & $+0.328$                   & $+0.715$       \\
$\rho(\lambda_+'',\lambda_0)$    & $-0.439$                   & $-0.695$       \\
$I(K^0_{e3})$                    & 0.154566(292)	      & 0.154354(389)  \\
$I(K^\pm_{e3})$                  & 0.158918(300)	      & 0.158700(400)  \\
$I(K^0_{\mu3})$                  & 0.102123(312)	      & 0.101643(424)  \\
$I(K^\pm_{\mu3})$                & 0.105073(321)	      & 0.104578(437)  \\
$\rho(I_{e3},I_{\mu3})$          & $+0.63$                    & $+0.89$        \\
\hline\hline
\end{tabular}
\caption{Averages of quadratic fit results for $K_{e3}$ and $K_{\mu3}$ slopes.}
\label{tab:l3ff}
}

The value of $\chi^2/{\rm ndf}$ for all measurements is terrible;
we are forced to quote the results with scaled errors. This leads to
errors on the phase-space integrals that are $\sim$60\% larger after inclusion
of the new  $K_{\mu3}$  NA48 data.

We have checked to see if the NA48 $K_{\mu3}$
data might show good consistency with the results of some other
experiment in a less inclusive average.
Fitting to only the $K_{\mu3}$ results from KTeV, NA48, and ISTRA+ gives
$\chi^2/{\rm ndf} = 27.5/6$ (0.01\%).
Fitting to only the $K_{L\:\mu3}$ results from KTeV, NA48 gives
$\chi^2/{\rm ndf} = 11.6/3$ (0.89\%).
The consistency of the NA48 data with these other measurements appears
to be poor in any case.

The evaluations of the phase-space integrals for all four modes are listed in each case.
Correlations are fully accounted for, both in the fits and in the evaluation
of the integrals. The correlation matrices for the integrals are of the
form
\begin{displaymath}
\begin{array}{cccc}
+1 & +1 & \rho & \rho \\
+1 & +1 & \rho & \rho \\
\rho & \rho & +1 & +1 \\
\rho & \rho & +1 & +1 \\
\end{array}
\end{displaymath}
where the order of the rows and columns is $K^0_{e3}$, $K^\pm_{e3}$,
$K^0_{\mu3}$, $K^\pm_{\mu3}$,
and $\rho = \rho(I_{e3},I_{\mu3})$ as listed in the table.

Adding the $K_{\mu3}$ data to the fit does not cause drastic changes
to the values of the phase-space integrals for the $K_{e3}$ modes:
the values for $I(K^0_{e3})$ and $I(K^\pm_{e3})$ in \Tab{tab:l3ff}
are qualitatively in agreement with those in \Tab{tab:e3ff}.
As in the case of the fits to the $K_{e3}$ data only, the significance of the
quadratic term in the vector form factor is strong ($3.6\sigma$ from the
fit to all data).

\section{Physics Results}
\label{sec:results}

\subsection{Determination of $f_{+}(0)V_{us}$ and $V_{us}/V_{ud}\times f_K/f_\pi$  }
 This section describe the results that are independent on the 
 theoretical parameters   $f_{+}(0)$ and $f_K/f_\pi$.

\subsubsection{Determination of $f_{+}(0)V_{us}$ }
 The value of $f_{+}(0)V_{us}$ has been determined from \ref{eq:Mkl3} using
 the world average values reported in \ref{sec:data}
 for lifetime, branching ratios and phase space integrals, and
 the radiative and SU(2) breaking corrections
 discussed in section~\ref{sec:master}.

 The results  are given in Table~\ref{tab:Vusf0},
 and are shown in \Fig{fig:Vusf0} for  $K_L \to \pi e \nu$, $K_L \to \pi \mu \nu$,
 $K_S \to \pi e \nu$, $K^\pm \to \pi e \nu$, $K^\pm \to \pi \mu \nu$, and for the combination.
\begin{table}[h]
\begin{center}
\begin{tabular}{l|c|c|c|c|c|c}
mode               & $f_{+}(0)V_{us}$  & \% err & BR   & $\tau$ & $\Delta$& Int \\
\hline
$K_L \to \pi e \nu$     & 0.21625(60)   & 0.28   & 0.09 & 0.19   & 0.15  & 0.09\\
$K_L \to \pi \mu \nu$   & 0.21675(66)   & 0.31   & 0.10 & 0.18   & 0.15  & 0.15\\
$K_S \to \pi e \nu$     & 0.21542(134)  & 0.67   & 0.65 & 0.03   & 0.15  & 0.09\\
$K^\pm \to \pi e \nu$   & 0.21728(84)   & 0.39   & 0.26 & 0.09   & 0.26  & 0.09\\
$K^\pm \to \pi \mu \nu$ & 0.21758(111)  & 0.51   & 0.40 & 0.09   & 0.26  & 0.15\\
 average & 0.21661(46)  &    &  &   &   & \\

\end{tabular}
\end{center}
\vskip 0.3cm
\caption{\label{tab:Vusf0}
 Summary of $f_{+}(0)V_{us}$ determination from all channels.}
\end{table}
\FIGURE{\figb f0vus;6;
\caption{Display of $f_{+}(0)V_{us}$ for all channels.}
\label{fig:Vusf0}}

 The average,   $\vert V_{us}\vert f_+(0)=0.21661(46)$, has an uncertainty of about of $0.2\%$.
  The results from the five modes are in good agreement, the
  fit probability is 58\%. 
 In particular, comparing the values of $f_{+}(0)V_{us}$
 obtained from $K^0_{\ell3}$ and $K^\pm_{\ell3}$ we obtain
 a value of the SU(2) breaking correction $$\delta^K_{SU(2)_{exp.}}= 2.86(39)\%$$ in agreement
 with the CHPT calculation reported in table~\ref{tab:iso-brk} $\delta^K_{SU(2)}= 2.36(22)\%$.

\subsubsection{Determination of $V_{us}/V_{ud}\times f_K/f_\pi$ }
\label{sec:fkfpvusvud}
Another determination of \Vus\ is obtained from $K_{\ell2}$ decays. 
The most important mode is  $K^+\to\mu^+\nu$ which has been recently 
updated by KLOE, so that the relative uncertainty is now about $0.3\%$.  
To reduce  hadronic uncertainties, in eq.~(\ref{eq:Mkl2})
 we have introduced the ratio 
$\Gamma(K^+\to\mu^+\nu)/\Gamma(\pi^+\to\mu^+\nu)$.

 Using the world average values
 of BR($K^\pm\to\mu^\pm\nu$) and of $\tau^\pm$ given in section\ref{sec:data}
 and the value of $\Gamma(\pi^\pm\to\mu^\pm\nu)=38.408(7)\mu s^{-1}$
 from \cite{bib:pdg} we obtain:
\[
V_{us}/V_{ud}\x f_K/f_\pi = 0.27599 \pm  0.00059
\]

\subsection{The parameters  $f_+$ and $f_K/f_\pi$ }

 For the time being, such a highly precise measurement could not be translated
 to a similar error on  the $\vert V_{us}\vert $ determination and therefore to
 the test of CKM unitarity.
 The obstacle is obviously the difficulty to keep 
 the theoretical uncertainties in  $f_+(0)$ and $f_K/f_\pi$ at the per-mil  level.
\subsubsection{$f_+(0)$ determination}
\label{sec:fofzero}
 In eq. (\ref{eq:Mkl3}), $\fp$  is defined in the absence
 of electromagnetic corrections and I-spin breaking effects.  While QCD is flavor blind the mass differences $m_u\ne m_d\ne m_s$ results in $\fp$ being different from unity and also being different for charged and neutral kaons.  In the following, by common convention, $\fp$ is refers to neutral kaons, the difference for charged kaons is accounted for in the term $\delta^{K}_{SU(2)}$ of Eq. (\ref{eq:Mkl3}). $\fp$ is calculable in non-perturbative QCD. 
In the flavor SU(3) limit, CVC ensures $\fp$=1, but then $m_K=m_\pi$ and there are no weak decays. 
We can write
\vspace*{-0.15cm}
 \be\label{chpt4}
\vspace*{-0.15cm}
 f_+(0)= 1 + f_2 + f_4 + \ldots
 \ee
In chiral perturbation theory (ChPT)  $f_2$ and $f_4$ stand for the leading and 
next-to-leading chiral corrections.  The Ademollo--Gatto theorem ensures that terms $\propto 
(m_s-m_u)$ are absent for vector transitions and $f_2=-0.023$  is unambiguously predicted in ChPT.   
The calculation of the chiral loop contribution, $\Delta(\mu)$ in
\be 
\renewcommand{\arraystretch}{0.5} 
f_4 =
\Delta(\mu) + f_4\vert^{\rm loc}(\mu)\,, 
\label{eq:f4ch}
\ee 
has been recently completed in ref.~\cite{tal}, but the full determination of $f_4$ 
necessitates an accurate estimation of the local counter-term 
 $f_4\vert^{\rm loc}(\mu)$, which is ${\cal O}(p^6)$.
\begin{figure}[htb]
\centering
\includegraphics[width=0.5\textwidth]{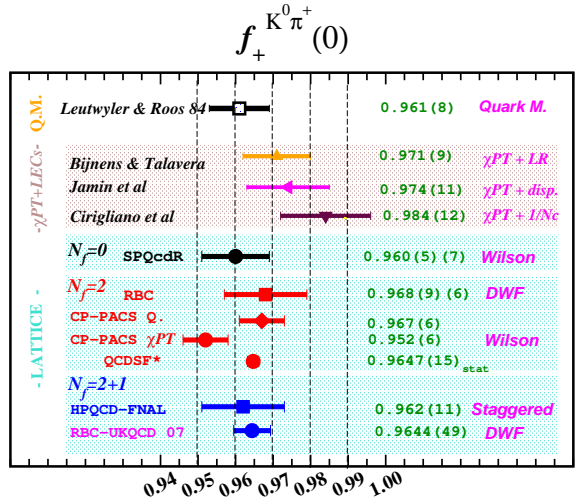}
\caption{\label{f0}Present determinations 
of $f_+(0)\equiv f_+^{K^0\pi^-}(0)$~\cite{LR,rbcf0,milcf0,f0} from lattice
QCD and other approaches. Hints from ChPt are exploited
 from all these approaches. }\end{figure}
\begin{figure}[htb]
\centering
\includegraphics[width=0.5\textwidth]{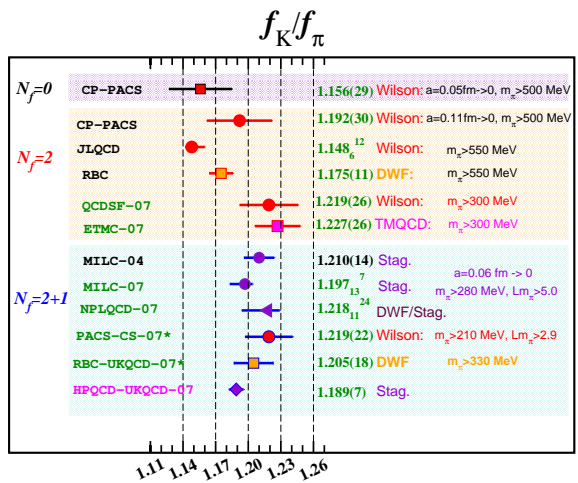}
\caption{\label{fk}Summary of $f_K/f_\pi$
estimates~\cite{milcfk,davies,rbcfk,fk}. All values are from Lattice QCD.               
In recent studies, sea quarks are getting light and data are  
matched to ChPt fits to determine the
Low-energy-Constants (LEC). 
}\end{figure}
Many theoretical approaches have been attempted 
over the years~\cite{f0}, essentially confirming the original
estimate by Leutwyler and Roos which was obtained in 
a simple quark model~\cite{LR}. The benefit of these new results, obtained using 
more sophisticated approaches, lies in the fact that we are nowadays in the position to control the 
systematic uncertainties of our calculations while with the quark models
this is not possible.  To stress the importance of the accurate determination of $f_4$, we 
should remind the reader that the experimental error on $\vert V_{us}\vert f_+(0)$ is only $0.2\%$,
 whereas the spread of
 theoretical estimates of $\fp$ is still at the $1\% \div2\%$ which is unsatisfactory.

Recent progress in lattice QCD gives us more optimism as far as the prospects
 of reducing the error on $\fp$ to well 
below $1\%$ are concerned~\cite{kan}.  Most of the currently available results
 obtained by using  lattice  QCD  worked with the ``heavy pions".
  Nevertheless, we can already see that the
 lattice QCD results are  systematically lower
 than those obtained by the ChPT-inspired models. 
An important step to resolving this issue has been recently made by 
the UKQCD-RBC collaboration~\cite{rbcf0}. 
Their preliminary  result $\fp=0.964(5)$
is obtained from the unquenched study  with 
$N_F=2+1$ flavors of the quarks which have a good chiral properties 
on the lattice with finite lattice 
spacing (so called, domain wall quarks), 
and their pions ($\gtrsim 300\,\mev$) are much lighter than what is reported in the previous 
lattice QCD studies. Their 
overall error is estimated to be  $~0.5\%$, which is very encouraging. 
It is important to emphasize that they observe the chiral logarithmic corrections, 
those which appear in the form factor as $f_2$, which was never observed 
before (most probably because other studies were restrained to heavier pions). 
One should keep in mind, however, that present study is performed at 
a single value of the lattice spacing (i.e. $a=0.12$ fm) and in a relatively small lattice
box.

\subsubsection{$f_K/f_\pi$ determination}

As we can see in eq.~(\ref{eq:Mkl2}), the QCD uncertainty enters with
 \be\label{chpt3}
 f_K/f_\pi= 1 + r_2 + \ldots
 \ee
In contrast to the semileptonic case discussed above,  the Ademollo--Gatto theorem does not
 apply in this case and 
 $r_2$  is not predicted unambiguously in ChPT. Rather one should fix the low energy constants from the lattice QCD 
 studies of   $  f_K/f_\pi$~\cite{milcfk}-\cite{fk}.  Such  obtained values are summarized 
in fig.~\ref{fk} from which we deduce that the  present  overall accuracy is about $1\%$. 
The  novelty are the new  lattice results with 
 $N_F=2+1$ dynamical quarks  and  pions as light as  $280$~MeV~\cite{milcfk,davies},
 obtained by using the so-called staggered quarks,
in which they covered a broad range of lattice spacings (i.e $a$=0.06 and 0.15 fm) and  
kept sufficiently large physical volumes (i.e. $m_\pi L\gtrsim 5.0$). 
It should be stressed, however, that the sensitivity of 
$f_K/f_\pi$ to the lighter pions is larger 
than in the computation of $\fp$ and that the chiral extrapolations are far more 
demanding in this case. Further improvement is expected soon.
   PACS-CS Collaboration~\cite{jam} has recently presented preliminary 
   results for $N_F=2+1$  clover quarks with pion masses 
   $\gtrsim 200$ MeV ($m_\pi L\gtrsim 2.9$ and $a=0.09$ fm). For the moment, 
   these simulations are on
   a single value of lattice volume and
   finite volume effects can be large. 
   
\subsubsection{A test of lattice calculation: the Callan-Treiman relation}
\label{sec:CTtest}
 As described in \ref{sec:ffpara} the Callan-Treiman relation fixes the
 value of scalar FF at $t=\Delta_{K\pi}$ (the so-called Callan-Trieman point) to the
 ratio of the pseudoscalar decay constants $f_K/f_\pi$ and
 the recent parametrization for the scalar FF\cite{stern} 
 allows this constraint to be exploited.
 The net result is that the  ratio $(f_K/f_\pi)/f_+(0)$ can be determined
 from the measurement of the scalar form factor.

 Very recentely KLOE~\cite{KLOE+07:m3FF} and NA48~\cite{NA48+06:m3FF} have presented results on the
 scalar form factor slopes using the dispersive parematrization.
 The older results of KTeV and ISTRA+ measurement of the scalar
 form factor slope performed using the 1st order Taylor expansion
 parametrization can be translated in the dispersive parametrization.
 The results are given in Table~\ref{tab:ffdis} for all 4 experiments in the case
 of pole parametrization for the vector form factor.
 The original KLOE and NA48 results are also shown for comparison.
\begin{table}[h]
\centering
\begin{tabular}{l|c|c}
Experiment & $\log(C)$ direct & $\log(C)^\dag$\\
\hline
KTeV   &           & 0.203(15)\\
KLOE   & 0.207(24) & 0.207(23)   \\
NA48   & 0.144(14) & 0.144(13)   \\
ISTRA+ &           & 0.226(13)   \\
\hline
\end{tabular}\\[1mm]
$^\dag$ {\small Estimated from $\lambda_0$ published.}\kern2cm\\
\caption{\label{tab:ffdis}
Experimental results of for log(C) and $\lambda_+$}
\end{table}

Figure \ref{fig:CTtest} shows the values for $f_+(0)$ determined from the scalar form factor slope
measurements obtained using the Callan-Treiman relation and
$f_K/f_\pi=1.189(7)$. The value of $f_+(0)=0.964(5)$ from UKQCD/RBC is also shown.
\FIGURE{\figb cttest;6;
\caption{Values for $f_+(0)$ determined from the scalar form factor slope using the Callan-Treiman relation and  $f_K/f_\pi=1.189(7)$. The UKQCD/RBC result $f_+(0)=0.964(5)$ is also
shown.\\[-1mm]}
\label{fig:CTtest}}
As already noticed in \ref{sec:data} the NA48 result is 
difficult to accommodate, and once compared with theory
it violates the Fubini-Furlan theorem $f_+(0)<1$.
 For this reason the NA48 result will be excluded when using the
Callan-Treiman constraint.

 The average of the experimental results on the FF's with the
 pole parametrization for the vector FF and the dispersive parematrization
 for the scalar form factor gives:
\be\eqalign{
  \lambda_+^c&=0.0256\pm 0.0002\cr
  \lambda_0^c&=0.0149\pm 0.0007\cr}
\ee
with correlation coefficient \minus0.32. 
 The above result is then 
 combined to the lattice determination of  $f_K/f_\pi=1.189(7)$ and 
 $f_+(0)=0.964(5)$ using the constraint given from the Callan-Treiman
 relation. The pole parematerization has been used to describe the
 vector form factor.
The results of the combination are given in table \ref{tab:ffsct},
\begin{table}[h]
\begin{center}
\begin{tabular}{c|c|c|c}
$\lambda_+^c$ & $\lambda_0^c$ & $f_+(0)$ & $f_K/f_\pi$ \\
\hline
  0.02563(20) & 0.01455(51) & 0.963(44) & 1.1913(61)\\
\hline
\multicolumn{4}{c}{correlation matrix}\\
\hline
          1.    & -0.24  &  0.11 & -0.13\\
                &  1.    & -0.45 &  0.54\\
                &        &  1.   &  0.27\\
                &        &       &  1.\\
\end{tabular}
\end{center}
\caption{\label{tab:ffsct}Results from the form factor fit.}
\end{table}

 where $\log C= \lambda_0^c \Delta_{K\pi}/m^2_\pi + 0.0398 \pm 0.0041$.
 
 The fit probability is 39\% confirming the agreement between
 experimental measurements and lattice determination.
 The accuracy of $f_K/f_\pi$ is slightly improved. The 
 improvement is better seen in the ratio 
 $f_+(0)/(f_K/f_\pi)$, directly related to the
 Callan-Treiman constraint. This latter improvement
 is very effective in the constraining scalar currents.
 
\subsection{Test of Cabibbo Universality or CKM unitarity}
To test CKM unitarity we use the value $f_{+}(0)V_{us}=0.21661(46)$ given in 
Table~\ref{tab:Vusf0}, $V_{us}/V_{ud}f_K/f_\pi = 0.27599(59)$ (see \ref{sec:fkfpvusvud}),
$f_+(0) = 0.964(5)$, and $f_K/f_\pi = 1.189(7)$. The results are:
\bea
 V_{us}&=& 0.22461\pm  0.00124\\
 V_{us}/V_{ud}&=& 0.23211\pm  0.00145
\eea
 These determinations can be used in a fit together with the 
 the recent evaluation of \Vud\ from
 $0^+\to0^+$ nuclear beta decays.
 The fit gives $\Vud = 0.97417(26)$ and $\Vus = 0.2253(9)$, 
 with $\chi^2/{\rm ndf} = 0.65/1$ (42\%). 
The unitarity constraint can also be included, in which 
case the fit gives $V_{us}=\sin\,\theta_C=\lambda=0.2255(7)$ and
$\chi^2/{\rm ndf}=0.80/2$ (67\%).
Both results are illustrated in \Fig{fig:vusuni}.
\FIGURE{\figb allfit;6;
\caption{Results of fits to \Vud, \Vus, and $\Vus/\Vud$.}\label{fig:vusuni}}
%
%
 As described in the introduction the test of CKM unitarity
 can be also interpreted as a test of universality of
 the lepton and quark gauge coupling:
\begin{equation}\eqalign{
G^2_{CKM}&=G^2_\mu\kern2mm\mbox{or}\kern2mm|V_{ud}|^2+|V_{us}|^2+|V_{ub}|^2\cr
&=1\ \mbox{and}\ G_F\equiv G_\mu.}
\label{eq:unitarity2}
\end{equation}
 Using the results of the fit we obtain:
\begin{equation}
G_{CKM} = (1.16624 \pm  0.00039)\times 10^{-5}\  {\rm GeV}^{-2}
\end{equation}
 In perfect agreement with the value obtained from the measurement
 of the muon lifetime:
\begin{equation}
G_{\mu} = (1.166371 \pm  0.000007)\times 10^{-5}\  {\rm GeV}^{-2}
\end{equation}
 The current accuracy of the lepton-quark universality,
 sets important constraint to the model builders of the beyond 
 SM physics scenarios. 
 For example, the presence of  a $Z^\prime$ (see Fig. \ref{fig:zpgraph}, left) would 
 affect the universality:
\begin{equation}
G_\mu=G_{CKM}
 \left[ 1-0.007
 Q_{eL}(Q_{{\mu}L}-Q_{dL}) \frac{ 2 \ln(m_{Z'}/m_W)}{m_{Z'}^2/m_W^2-1}\right]
\label{eq:zprimo}
\end{equation}
\FIGURE{\figb zph;9;
\caption{\label{fig:zpgraph} Z' and Higgs exchange.}}

 In case of $Z^\prime$ in SO(10) grand unification theories,
 ($Q_{eL}=Q_{{\mu}L}=-3Q_{dL}=1$) we obtain
  $m_{Z^\prime}>700$~GeV at 95\% CL, 
 to be compared 
 the one set through the direct collider searches,
 $m_{Z^\prime}>720$~GeV~\cite{bib:pdg}.

\subsubsection{Bounds from helicity suppressed processes}
 A particularly interesting observable is the ratio of the $V_{us}$ values obtained
 from helicity suppressed processes to that obtained from helicity allowed modes:
 $R_{l23}=V_{us}(K_{\ell 2})/V_{us}(K_{\ell 3})$. 
 According to\cite{gino,stern} the value of $R_{l23}$ would be affected by the
 presence of scalar density or extra right-handed currents:
\be
 R_{l23} = 1 + \delta R_{l23}
\ee
 To improve the accuracy of the determination of $R_{l23}$ we use the 
 the values of $f_+(0)$ and $f_K/f_\pi$ obtained in \ref{sec:CTtest}
\footnote{ $R_{l23}$ depends only on the ratio $f_+(0)$/($f_K/f_\pi$)}.
 In addition, in this scenario both $0^+\to0^+$ nuclear beta decays
 and $K_{\ell3}$ are not affected and the unitarity constraint for these modes
 can be used. The fit described in the previous section
 has been performed assuming unitarity and allowing for
 two different values of $V_{us}$ from helicity suppressed($K\to \mu \nu$) and
 allowed modes ($K\to \ell \pi \nu$, $0^+\to0^+$ nuclear beta decays).
 We obtain: 
\be
R_{l23} =  1.0028 \pm  0.0059
\ee
If the scalar current is due to charged Higgs exchange as shown in 
Fig. \ref{fig:zpgraph}, right:
\be
 \delta R_{l23}  =
(1-tan^2 \beta \frac{m_{K^\pm}^2/m_{H^\pm}^2}{1+0.01\tan \beta})
\ee
 and the measurement of $ R_{l23}$ can be used to set bounds
 on the charged Higgs mass and $\tan\beta$.
 Figure \ref{fig:higgskmunu} shows the exlcuded region at 95\%
 CL in the charged Higgs mass-$\tan\beta$ plane.
\FIGURE{\figb higgskmnu;6;
\caption{\label{fig:higgskmunu}Exlucluded region in the charged Higgs mass-$\tan\beta$ plane.
The region excluded by  $B\to \tau \nu $ is also indicated.}}
 The measurement of  BR($B \to \tau \nu$)\cite{bib:btaunu}
 can be also used to set bound on the charged
 Higgs mass-$\tan\beta$ palne. While the $B\to\tau \nu$
 can exclude quite an extensive region of this plane,
 there is an uncovered region in the exclusion corresponding 
 to the change of sign of the correction.
 This region is fully covered by the $K\to \mu \nu$ and
 the positive correction solution for the $B\to\tau \nu$
 is fully excluded.

\subsection{Test of Lepton Flavour violation}
\subsubsection{Lepton universality and \boldmath{$K_{\ell 3}$} decays}
Search for lepton flavour violation (LFV) in the semileptonic decays
 $K_{e3}$ and $K_{\mu3}$
is a test of the vector current of the weak interaction. It can therefore be compared
to LFV tests in $\tau$ decays, but is different to LFV searches in
$\pi_{l2}$ and $K_{l2}$ decays.

The results on the parameter
 $r_{\mu e} = 
R_{K_{\mu3}/K_{e3}}^{\rm{Exp}}/R_{K_{\mu3}/K_{e3}}^{\rm{SM}}$ is
\be
r_{\mu e} = 1.0040 \pm 0.0044,
\ee
in excellent agreement with lepton universality. 
Furthermore, with a precision of $0.5\%$ the test in $K_{l3}$ decays 
has now reached the sensitivity of $\tau$ decays.
The accuracy of this test slightely improves
 using FF's slope values of section\ref{sec:CTtest} 
\be
r_{\mu e}= 0.9998 \pm  0.0040
\ee

\subsubsection{Lepton universality tests in \boldmath{$K_{\ell 2}$} decays}
The ratio $R_K = \Gamma(K_{e2})/\Gamma(K_{\mu2})$ can be precisely calculated within the Standard Model.
 Neglecting radiative corrections, it is given by 
\be
R_K^{(0)} = \frac{m_e^2}{m_\mu^2} \: \frac{(m_K^2 - m_e^2)^2}{(m_K^2 - m_\mu^2)^2} = 2.569 \times 10^{-5},
\ee
and reflects the strong helicity suppression of the electronic channel.
Radiative corrections have been computed within the model of vector meson
 dominance~\cite{ciriglianokl2},
yielding a corrected ratio of
\be
R_K = R_K^{(0)} ( 1 + \delta R_K^{\text{rad.corr.}}) = 2.569 \times 10^{-5} \times ( 0.9622 \pm 0.0004 ) =(2.477 \pm 0.001) \times 10^{-5}. 
\ee

Because of the helicity suppression of $K_{e2}$ in the SM, the decay amplitude is a
 prominent candidate
for possible sizeable contributions from new physics beyond the SM. Moreover,
 when normalizing to $K_{\mu2}$ decays, it is one of the few kaon decays, for which
 the SM-rate is predicted with very high accuracy.
Any significant experimental deviation from the prediction would immediately be
 evidence for new physics.
 However, this new physics would need to violate lepton universality
 to be visible in the ratio $K_{e2}/K_{\mu2}$.

 Recently it has been pointed out, that in a SUSY framework sizeable violations of
 lepton universality can be expected
 in $K_{l2}$ decays~\cite{paride}. At tree level, 
 lepton flavour violating terms are forbidden in the MSSM. 
 Loop diagrams, however, should induce lepton flavour violating Yukawa couplings
 as  $H^+ \to l \nu_\tau$ to the charged Higgs boson $H^+$.
 Making use of this Yukawa coupling, the dominant non-SM contribution
 to $R_K$ modify the ratio to
\be
R_K^{\text{LFV}} \approx R_K^{\text{SM}} \left[ 1 + \left( \frac{m_K^4}{M_{H^\pm}^4} \right) \left( \frac{m_\tau^2}{M_e^2} \right) |\Delta_{13}|^2 \tan^6 \beta \right].
\label{eqn:susy}
\ee
 The lepton flavour violating term $\Delta_{13}$ should be of the order 
 of $10^{-4}-10^{-3}$, as expected from neutrino mixing. 
 For moderately large $\tan \beta$ and $M_{H^\pm}$, SUSY contributions may therefore
  enhance $R_K$ might by up to a few percent.
 Since the additional term in Eqn.~\ref{eqn:susy} goes with the forth power of the
 meson mass, no similar effect is expected in $\pi_{l2}$ decays.
\FIGURE{\figb rklim;6;\caption{Exclusion limits at $95\%$ CL on $\tan \beta$ and the charged
Higgs mass $M_{H^\pm}$ from $\vert V_{us}\vert_{K\ell2}/\vert V_{us}\vert_{K\ell3}$ for different
values of $\Delta_{13}$.}\label{fig:susylimit}}

 The world average result for $R_K$\ref{sec:data}
 gives strong constraints for $\tan \beta$ and $M_{H^\pm}$ (Fig.~\ref{fig:susylimit} (left)).
 For a moderate value of $\Delta_{13} \approx 5 \times 10^{-4}$, $\tan \beta > 50$
 is excluded for charged Higgs masses up to 1000~GeV/$c^2$ at 95\% CL.

\section{Conclusions}
Many new precise mesaurements about $K_{\ell3}$ and $K_{\ell2}$ decays
properties have been performed recently allowing precise tests
 of the Standard Model to be performed. We determine:
\[
f_{+}(0)\times V_{us} = 0.21661(46)
\]
\[
f_K/f_\pi \times V_{us}/V_{ud} = 0.27599(59) 
\]
using $f_K/f_\pi=1.189(7)$ and  $f_+(0)=0.964(5)$
from recent lattice evaluation we obtain:
\bea
 V_{us}&=& 0.2246(12)\\
 V_{us}/V_{ud}&=& 0.2321(15)
\eea
in very good agreement with $\Vud = 0.97417(26)$ from
$0^+\to0^+$ nuclear beta decays.
These determinations can be used to evaluate
the Fermi constant from the quark sector:
 \[
G_{CKM} = (1.16624 \pm  0.00039)\times 10^{-5}\  {\rm GeV}^{-2}
\]
to be compared with that obtained from the muone lifetime:
\[
G_{\mu} = (1.166371 \pm  0.000007)\times 10^{-5}\  {\rm GeV}^{-2}
\]
 The improved accuracy of these measurements sets non-trivial constraints
 for physics beyond the Standard Model. In particular, the comparison
 of the \Vus\ determinations from helicity suppressed and allowed
 processes sets a competitive lower bound on the charged higgs mass
 as a function of $\tan \beta$.

 Very recentely the test of lepton flavour violation
 in helicity suppressed processes through the
 observable $R_K = \Gamma(K_{e2})/\Gamma(K_{\mu2})$
 as been improved significantly. The new world
 average
\[
R_K = \BR{K_{e2}}/\BR{K_{\mu2}}= (2.45 \pm 0.11) \times 10^{-5}
\]
 gives the best constraint for the $e-\tau$ LFV coupling of
 MSSM.

\begin{acknowledgments}
I would like thank all the members of the FlaviaNet Kaon Working
Group who contributed significantly to this work:
 Paolo Franzini, Gino Isidori, Federico Mescia,  Matthew Moulson, and Matteo Palutan.
 I also thank Juliet Lee Franzini for the careful reading of this report.
This work is  supported in part by the 
EU contract No. MTRN-CT-2006-035482(FlaviaNet).
\end{acknowledgments}




\begin{thebibliography}{99}




\bibitem{flavia}
FlaviaNet Kaon working group: \verb"http://www.lnf.infn.it/wg/vus/"

\bibitem{mulan} MuLan Collaboration, D.B.~Chitwood \etal, \emph{Phys.\
  Rev.\ Lett.}\ {\bf 99} (2007) 032001.   
  
\bibitem{t&h} J.C. Hardy and I.S. Towner, arXiv:0710.3181v1 [nucl-th].

\bibitem{Sirlin:1981ie}
A.~Sirlin,
Nucl.\ Phys.\ B {\bf 196}, 83 (1982).
\bibitem{ciriglianokl2}W.~J.~Marciano,
  Phys.\ Rev.\ Lett.\  {\bf 93}, 231803 (2004)
  [arXiv:hep-ph/0402299]; V.~Cirigliano and I.~Rosell,
  JHEP {\bf 0710}, 005 (2007)
  [arXiv:0707.4464 [hep-ph]]; arXiv:0707.3439 [hep-ph].

\bibitem{Cirigliano}
V.~Cirigliano  {\it et al.},
Eur.\ Phys.\ J.\ C {\bf 35} (2004) 53;
Eur.\ Phys.\ J.\ C {\bf 23} (2002) 121;
T.~C.~Andre,
hep-ph/0406006.
\bibitem{neufeld} V.~Cirigliano, H.~Neufeld and I.~Rosell, work in preparation.


\bibitem{stern}
  V.~Bernard, M.~Oertel, E.~Passemar and J.~Stern,
  arXiv:0707.4194 [hep-ph];
  Phys.\ Lett.\  B {\bf 638}, 480 (2006)
  [arXiv:hep-ph/0603202].

\bibitem{leut}H.~Leutwyler private comunication.



\bibitem{KTeV+04:BR}T.~Alexopoulos {\it et al.}  [KTeV Collaboration],
  Phys.\ Rev.\  D {\bf 70}, 092006 (2004)
  [arXiv:hep-ex/0406002].

\bibitem{NA48+04:BR}
  A.~Lai {\it et al.}  [NA48 Collaboration],
  Phys.\ Lett.\  B {\bf 602}, 41 (2004)
  [arXiv:hep-ex/0410059].

\bibitem{Lit04:ICHEP}
  L.~Litov  [NA48 Collaboration],
  arXiv:hep-ex/0501048.
\bibitem{KLOE+05:tauL}
  F.~Ambrosino {\it et al.}  [KLOE Collaboration],
  Phys.\ Lett.\  B {\bf 626}, 15 (2005)
  [arXiv:hep-ex/0507088].
\bibitem{KLOE+06:BR}
  F.~Ambrosino {\it et al.}  [KLOE Collaboration],
  Phys.\ Lett.\  B {\bf 632}, 43 (2006)
  [arXiv:hep-ex/0508027].
\bibitem{KLOE+06:KLpp}
  F.~Ambrosino {\it et al.}  [KLOE Collaboration],
  Phys.\ Lett.\  B {\bf 638}, 140 (2006)
  [arXiv:hep-ex/0603041].
\bibitem{NA48+06:KLpp}
  A.~Lai {\it et al.}  [NA48 Collaboration],
  Phys.\ Lett.\  B {\bf 645}, 26 (2007)
  [arXiv:hep-ex/0611052].
\bibitem{KLOE+06:KSe3}
  F.~Ambrosino {\it et al.}  [KLOE Collaboration],
  Phys.\ Lett.\  B {\bf 636}, 173 (2006)
  [arXiv:hep-ex/0601026].
\bibitem{KLOE+06:KSpp}
  F.~Ambrosino {\it et al.}  [KLOE Collaboration],
  Eur.\ Phys.\ J.\  C {\bf 48}, 767 (2006)
  [arXiv:hep-ex/0601025].
\bibitem{NA48:KSe3}
 J.~R.~Batley {\it et al.},
  Phys.\ Lett.\  B {\bf 653}, 145 (2007).

\bibitem{NA48+07:BR}
  J.~R.~Batley {\it et al.}  [NA48/2 Collaboration],
  Eur.\ Phys.\ J.\  C {\bf 50}, 329 (2007)
  [arXiv:hep-ex/0702015].
\bibitem{Rom06:ke3}
 V.~I.~Romanovsky {\it et al.},
  arXiv:0704.2052 [hep-ex].
\bibitem{KLOE:kl3pm}  F.~Ambrosino {\it et al.}  [KLOE Collaboration],
  arXiv:0707.2532 [hep-ex].
\bibitem{KLOE05:kl2}
  F.~Ambrosino {\it et al.}  [KLOE Collaboration],
  Phys.\ Lett.\  B {\bf 632}, 76 (2006)
  [arXiv:hep-ex/0509045].
\bibitem{KLOE:pipo} F.~Ambrosino {\it et al.}  [Kloe Collaboration],
  arXiv:0707.2654 [hep-ex].
\bibitem{KLOE:taupm} F.~Ambrosino {\it et al.}  [KLOE Collaboration],
    arXiv:0705.4408v3 [hep-ex].

\bibitem{bib:pdg} PDG, W.-M.~Yao \etal, \emph{J.\ Phys.} {\bf G33} (2006)
  1.  
\bibitem{bib:Ke2_2003} L.~Fiorini, PoS {\bf HEP2005}, 288 (2006); \\
                       L.~Fiorini, ph.D.~thesis, Pisa (2005);


\bibitem{bib:Ke2_2004} V.~Kozhuharov, KAON07 International Conference.

\bibitem{bib:Ke2_KLOE} F.~Ambrosino {\it et al.}  [Kloe Collaboration],
    arXiv:0707.4623v1 [hep-ex].

\bibitem{KTeV+04:FF}
  T.~Alexopoulos {\it et al.}  [KTeV Collaboration],
  Phys.\ Rev.\  D {\bf 70}, 092007 (2004)
  [arXiv:hep-ex/0406003].

\bibitem{KLOE+06:FF}
  F.~Ambrosino {\it et al.}  [KLOE Collaboration],
  Phys.\ Lett.\  B {\bf 636}, 166 (2006)
  [arXiv:hep-ex/0601038].
\bibitem{ISTRA+04:e3FF}
  O.~P.~Yushchenko {\it et al.},
  Phys.\ Lett.\  B {\bf 589}, 111 (2004)
  [arXiv:hep-ex/0404030].
\bibitem{NA48+04:e3FF}
  A.~Lai {\it et al.}  [NA48 Collaboration],
  Phys.\ Lett.\  B {\bf 604}, 1 (2004)
  [arXiv:hep-ex/0410065].
\bibitem{KLOE+07:m3FF}F.~Ambrosino {\it et al.}  [KLOE Collaboration],
  arXiv:0707.4631 [hep-ex].
\bibitem{ISTRA+04:m3FF}O. P. Yushchenko {\it et al.}, Phys.\ Lett.\ B 581 (2004) 31.
\bibitem{NA48+06:m3FF}A.~Lai {\it et al.} [NA48~Collaboration], Phys.\ Lett.\ B 647 (2007) 341.

\bibitem{tal}
J.~Bijnens and P.~Talavera,
Nucl.\ Phys.\ B {\bf 669}, 341 (2003); P.~Post and K.~Schilcher,
Eur.\ Phys.\ J.\ C {\bf 25}, 427 (2002).

\bibitem{LR} 
H.~Leutwyler and M.~Roos,
Z.\ Phys.\ C {\bf 25} (1984) 91.

\bibitem{rbcf0}
 D.~J.~Antonio {\it et al.},
  arXiv:hep-lat/0702026;

\bibitem{milcf0}
  M.~Okamoto  [Fermilab Lattice, MILC and HPQCD Collaborations],
  Int.\ J.\ Mod.\ Phys.\  A {\bf 20}, 3469 (2005);

\bibitem{f0}
  J.~Portoles,
  arXiv:hep-ph/0703093; M.~Jamin, J.~A.~Oller and A.~Pich,
JHEP {\bf 0402}, 047 (2004); V.~Cirigliano, G.~Ecker, M.~Eidemuller, R.~Kaiser, A.~Pich and J.~Portoles,
  JHEP {\bf 0504}, 006 (2005)
  [arXiv:hep-ph/0503108];
  N.~Tsutsui {\it et al.}  [JLQCD Collaboration],
  PoS {\bf LAT2005}, 357 (2006)
  [arXiv:hep-lat/0510068];
  C.~Dawson, T.~Izubuchi, T.~Kaneko, S.~Sasaki and A.~Soni,
  Phys.\ Rev.\  D {\bf 74}, 114502 (2006)
  [arXiv:hep-ph/0607162];
   D.~Becirevic {\it et al.},
  Nucl.\ Phys.\  B {\bf 705}, 339 (2005)
  [arXiv:hep-ph/0403217]; D.~Brommel {\it et al.},
  arXiv:0710.2100 [hep-lat].


\bibitem{milcfk}  C.~Bernard {\it et al.},
  arXiv:0710.1118 [hep-lat].

\bibitem{davies}
   E.~Follana, C.~T.~H.~Davies, G.~P.~Lepage and J.~Shigemitsu  [HPQCD
                  Collaboration],
  arXiv:0706.1726 [hep-lat].

\bibitem{rbcfk}
 C.~Allton {\it et al.}  [RBC and UKQCD Collaborations],
  Phys.\ Rev.\  D {\bf 76}, 014504 (2007)
  [arXiv:hep-lat/0701013].

\bibitem{fk}
S.~Aoki {\it et al.}  [CP-PACS Collaboration],
  Phys.\ Rev.\  D {\bf 67}, 034503 (2003)
  [arXiv:hep-lat/0206009]; A.~Ali Khan {\it et al.}  [CP-PACS Collaboration],
  Phys.\ Rev.\  D {\bf 65}, 054505 (2002)
  [Erratum-ibid.\  D {\bf 67}, 059901 (2003)]
  [arXiv:hep-lat/0105015];
S.~Aoki {\it et al.}  [JLQCD Collaboration],
  Phys.\ Rev.\  D {\bf 68}, 054502 (2003)
  [arXiv:hep-lat/0212039];
 Y.~Aoki {\it et al.},
  Phys.\ Rev.\  D {\bf 72}, 114505 (2005)
  [arXiv:hep-lat/0411006];
  M.~Gockeler {\it et al.},
  PoS {\bf LAT2006}, 160 (2006)
  [arXiv:hep-lat/0610071];
  C.~Aubin {\it et al.}  [MILC Collaboration],
  Phys.\ Rev.\  D {\bf 70}, 114501 (2004)
  [arXiv:hep-lat/0407028];S.~R.~Beane, P.~F.~Bedaque, K.~Orginos and M.~J.~Savage,
  Phys.\ Rev.\  D {\bf 75}, 094501 (2007)
  [arXiv:hep-lat/0606023]; T.~Ishikawa {\it et al.},
  PoS {\bf LAT2006}, 181 (2006)
  [arXiv:hep-lat/0610050];   B.~Blossier {\it et al.}  [European Twisted Mass Collaboration],
  arXiv:0709.4574 [hep-lat].

\bibitem{kan}
Juettner, plenary talk at Lattice 2007; T.~Kaneko,
  arXiv:0710.0698 [hep-ph].

\bibitem{jam}
 D.~Kadoh {\it et al.}  [CS Collaboration],
  arXiv:0710.3467 [hep-lat]; N.~Ukita {\it et al.}  [CS Collaboration],
  arXiv:0710.3462 [hep-lat].

\bibitem{gino}
  G.~Isidori and P.~Paradisi,
  Phys.\ Lett.\  B {\bf 639}, 499 (2006)
  [arXiv:hep-ph/0605012]; W.~S.~Hou,
  Phys.\ Rev.\  D {\bf 48}, 2342 (1993).

\bibitem{bib:btaunu} K.~Ikado {\it et al.},
  Phys.\ Rev.\ Lett.\  {\bf 97}, 251802 (2006)
  [arXiv:hep-ex/0604018];
  B.~Aubert {\it et al.}  [BABAR Collaboration],
  Phys.\ Rev.\  D {\bf 76}, 052002 (2007)
  [arXiv:0708.2260 [hep-ex]].


 \bibitem{paride}
A.~Masiero, P.~Paradisi and R.~Petronzio,
hep-ph/0511289.


\end{thebibliography}
\end{document}